\documentclass[a4paper,12pt,oneside,dvips,driver]{article}
\usepackage[english]{babel}
\usepackage{pstricks,pst-node,pst-coil}
\usepackage{epsfig}
\usepackage{amscd}
\usepackage{amssymb}
\usepackage[T1]{fontenc}
\usepackage{latexsym}
\usepackage[]{psfrag} 

\psfrag{el}{$\tilde{l}$}
\psfrag{eta0}{$\eta^{(0)}/(\sigma\xi_B)$}
\psfrag{tau1}{$\tau_1/\sigma$}

\newcommand{\beq}{\begin{equation}}
\newcommand{\eeq}{\end{equation}}
\newcommand{\bdm}{\begin{displaymath}}
\newcommand{\edm}{\end{displaymath}}

\begin{document}

\author{P. Jakubczyk  and  M. Napi\'{o}rkowski\\ Instytut Fizyki
Teoretycznej, Uniwersytet Warszawski  \\  00-681 Warszawa, Ho\.za  69,
Poland } 
\title{The influence of droplet size on line tension} 
\date{} 
\maketitle 

\begin{abstract}

{Within the effective interfacial Hamiltonian approach we evaluate 
the excess line free energy associated with cylinder-shaped droplets 
sessile on a stripe-like chemical inhomogeneity of a planar substrate. In the 
case of short-range intermolecular forces the droplet morphology and 
the corresponding expression for the line tension - which includes 
the inhomogeneity finite 
width effects - are derived and discussed as functions 
of temperature and increasing width. 
The width-dependent contributions to the line tension change their structure at the 
stripe wetting temperature $T_{W1}$:  for $T<T_{W1}$ they decay exponentially 
while for $T>T_{W1}$ the decay is algebraic. In addition, a geometric construction  
of the corresponding contact angle is  carried out and its implications are 
discussed. } \\


\noindent{\noindent Corresponding author: Pawe{\l} Jakubczyk, Instytut 
Fizyki Teoretycznej, Uniwersytet Warszawski 0-681 Warszawa, Ho\.za 69, 
Poland; tel. 0048225532317, e-mail: Pawel.Jakubczyk@fuw.edu.pl}
\end{abstract}

\newpage

\section{Introduction}
\renewcommand{\theequation}{1.\arabic{equation}} 
\setcounter{equation}{0}
\vspace*{0.5cm}

A lot of theoretical and experimental effort has recently been made towards 
understanding the adsorption on substrates equipped with geometric  
and 
chemical structure. This research exposed a variety of phenomena and 
associated problems, to mention  the filling transitions [1-4], position and 
droplet's size-dependent contact angles and interfacial morphology structures 
[5-10], 
discontinuous changes in droplet shapes as function of volume [11-12], or the 
influence of substrate's structure on the existence and order of 
wetting transitions as compared to the homogeneous and planar 
substrate case [13-17]. 
In many such systems the interfacial morphology leads to the existence of 
three-phase contact lines and  in consequence  to the 
line tension \cite{RowWid}. This quantity's properties, and especially its 
behaviour close to the  wetting transitions stimulated many 
discussions and controversies in recent years [19-29]. 

In this paper we are concerned with semi-infinite fluid in a thermodynamic state 
close to its bulk 
liquid-vapour coexistence, in the presence of a substrate consisting of a
single stripe-like inhomogeneity placed on an otherwise chemically homogeneous 
and planar solid surface, see Fig.1. The system is translationally invariant 
along the stripe, say in the $y$-direction,   
and the equilibrium adsorption morphology varies only in the $x$-direction 
perpendicular to the stripe, the width of which is denoted by $2R$. 
Thus the chemical inhomogeneity of the substrate imposes nonuniformity of 
the liquid-like layer adsorbed at the substrate. Its thickness varies in the 
$x$-direction and induces linear contribution 
to the  fluid free energy. This contribution evaluated per unit 
length of the stripe will be denoted $\eta$, and called the line 
tension. It is a function of  temperature $T$, the inhomogeneity width $2R$, 
and additionally depends on  
quantities characterizing both the stripe and the rest of the substrate. 
These additional quantities 
will be represented by the relevant wetting temperatures of two parts of the 
substrate. The $y=const $ section 
of the system can be looked upon as a two-dimensional droplet sessile on a 
one-dimensional substrate inhomogeneity, see Fig.2. We are particularly 
interested in contributions to the droplet line tension resulting from its 
finite size - their size and temperature dependence. 

In addition to analyzing the droplet morphology and the contributions 
to the line tension we perform geometric constructions of the relevant 
contact angles. These constructions correspond to different choices of 
the three phase contact line and lead to different predictions concerning the 
behaviour of the contact angle in the limit $R\to\infty$. In effect, only 
one of them turns out to have practical meaning.

This work is arranged as follows. In Section 2 we specify the model and 
discuss its limitations. A set of constrained equilibrium profiles 
parametrized by the droplet's height at its centre is constructed and the 
constrained excess line free energy expressions as functions of droplets' height,  
inhomogeneity width $2R$, temperature, and the corresponding wetting temperatures  
are derived. In Section 3 we construct the 
expansion for equilibrium droplet height in the limit of large $R$. We 
recover some of the 
earlier specific results \cite{Conf} together with new expressions for the 
corrections up to the order 
which is necessary to calculate the leading $R$-dependent contributions to 
the line tension. 
In Section 4 the expressions for the line tension are derived and 
discussed. In Section 5 we analyze the 
contact angles obtained from the equilibrium liquid - vapour interfacial shapes.

\section{The model}
\renewcommand{\theequation}{2.\arabic{equation}} 
\setcounter{equation}{0}
\vspace*{0.5cm}

We consider a planar, chemically inhomogeneous substrate remaining in 
contact with a fluid. The system's thermodynamic state is infinitisimally close 
to the fluid's bulk liquid-vapour coexistence, i.e., $\mu=\mu_{sat}^{-}$, where 
$\mu_{sat}$ denotes the chemical 
potential at liquid-vapour coexistence. The inhomogeneity has the form of 
a stripe of width $2R$. The substrate is translationally 
invariant in the $y$-direction defined by the inhomogeneity's orientation, 
see Fig.1. The wetting temperature of the stripe $T_{W1}$ is lower 
than that of the remaining part of the substrate  $T_{W2}$, i.e., 
$T_{W1}<T_{W2}$, 
so that the liquid phase is preferentially adsorbed on the stripe. The 
system's temperature $T$ is assumed lower than $T_{W2}$ so that the 
equilibrium thickness of the adsorbed layer $\bar{l}(x)$ converges in 
the limit $x\to\pm\infty$ to a finite value determined by the 
properties of substrate 2 alone, see Fig.2. 

\begin{figure}[h]
\begin{center}
\includegraphics[width=0.75\textwidth]{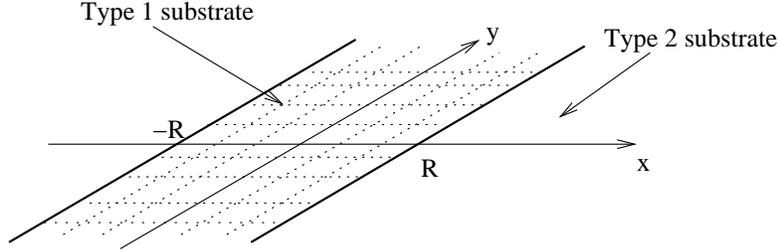}
\caption{A schematic illustration of a planar substrate 
containing a stripe-like inhomogeneity. The wetting temperature of 
the stripe ($T_{W1}$) is assumed lower than the substrate 2 wetting 
temperature ($T_{W2}$). The domain width ($2R$) is considered much 
larger than bulk correlation lengths.}   \end{center}
\end{figure}

\begin{figure}[h]
\begin{center}
\includegraphics[width=0.55\textwidth]{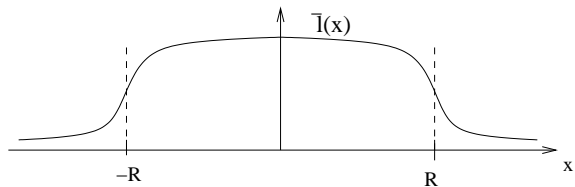}
\caption{Schematic plot of the $y=const$ section of the fluid density 
in the vicinity of the stripe, and the corresponding equilibrium 
shape of the liquid-vapor interfacial profile denoted by  $\bar{l}(x)$. } 
\end{center}
\end{figure}

A suitable description of this system, valid on length scales 
much larger than the bulk correlation length 
$\xi_B$ is provided by the interfacial Hamiltonian model 
\cite{Bau1}, \cite{Bau2} 
\beq
\label{H}
\mathcal{H}[l]=\int dx \int dy 
\, \Bigg[\frac{\sigma}{2}(\nabla l)^2+V(l,x)\Bigg] \quad. 
\eeq
Here $l(x,y)$ denotes the interface position, $\sigma$ - the liquid-vapour 
surface tension, and $V(l,x)$ - the 
effective interfacial potential. The borderline between substrates of 
types 1 and 2 is assumed to be of 
the order of $\xi_B$ and the crossover in $V(l,x)$ is anticipated to 
occur over a distance of the same 
magnitude for systems with short-ranged intermolecular forces 
\cite{Conf}, \cite{Bau1}. 
Consequently, we model $V(l,x)$ as
\beq
\label{V}
V(l,x)=\Theta (R-|x|)V_1(l)+\Theta (|x|-R) V_2(l),
\eeq
where $V_1(l)$ and $V_2(l)$ are the effective interfacial potentials 
corresponding to homogeneous substrates 
of type 1 and 2, respectively, and $\Theta (x)$ is the Heaviside step 
function. For short-range forces each of the potentials  
$V_i(l)$ ($i=1,2$) has the exponential form 
(see e.g., \cite{R1}, \cite{Conf})
\beq
\label{Vsh}
V_i(l)= \tau_i e^{-l/\xi_B}+b e^{-2 l/\xi_B}+... \quad,
\eeq
where $\tau_i=a(T-T_{Wi})$, $a$ being a positive constant, 
and  $\xi_B$ denotes the bulk correlation 
length in the adsorbed liquid phase. The positive parameter $b$ controls 
repulsion of the interface from the substrate at short distances. 
For simplicity, within our model, it has the same value for substrates 
of both kinds. \\
Thorough this work we apply the mean-field approach according to which 
the equilibrium grand cannonical average value of the adsorbed layer's 
thickness $<l(x)>$ is identified with $\bar{l}(x)$ which minimizes the 
Hamiltonian (\ref{H}).  One should also note that the full drumhead 
expression in the interfacial Hamiltonian (see e.g., \cite{Dobbs}) is 
substituted with the gradient term in (\ref{H}). 
This approximation is valid for small differences between both 
substrates which in turn is controlled by the difference 
$\tau_1-\tau_2$. The equilibrium profile $\bar{l}(x,y)$ is invariant 
with respect to translation in the $y$ direction 
and in addition $\bar{l}(x,y)=\bar{l}(x)=\bar{l}(-x)$. From now on the 
considered values of $x$ are limited to $x\in [0,\infty[$. 

Minimizing the functional (\ref{H}) leads to the Euler-Lagrange equation
\beq 
\label{EL}
\sigma\frac{d^2 {l}}{dx^2}=\frac{dV}{d{l}}
\eeq
together with the derivative continuity condition
\beq
\label{cp}
\left.\frac{d{l}}{dx}\right|_{x=R^-}=\left.\frac{d{l}}{dx}\right|_{x=R^+}\quad. 
\eeq
The minimization procedure is performed in two steps. First we solve 
Eq.(\ref{EL}) subject to the boundary conditions: 
$\frac{d{l}}{dx}|_{x=0}=0$, $\lim_{x\to\infty}{l}(x)=l_{\pi 2}$, 
where $l_{\pi 2}$ corresponds to the minimum of $V_2(l)$. 
The droplet's height at the centre $l_0={l}(x=0)$ is for the time 
being kept as fixed but arbitrary parameter. This way 
we obtain a set of constrained equilibrium profiles denoted by 
$\{\tilde{l}(x,l_{0})\}$; they are parametrized by $l_0$. In the 
second step the equilibrium droplet's height 
$\bar l_0$ and thus $\bar l (x)=\tilde l(x,\bar l_0)$ 
are determined by 
minimizing $H[\tilde{l}(x,l_0)]$ with respect to $l_0$. \\ 

For this purpose two cases must be distinguished: 1) $V_1(l_0)<0$; 
and 2)  $V_1(l_0)\geq 0$. 
In case 1), upon applying (\ref{V}), (\ref{Vsh}) we obtain the 
following formulae for the 
constrained equilibrium profiles corresponding to fixed value $l_0$:
\beq
\label{l1}
\frac{\tilde{l}(x,l_0)}{\xi_B} = \left\{ \begin{array}{ll}
\log\Big[\frac{1}{2V_1(l_0)}\Big(\tau_1+
(\tau_1+2be^{-l_0/\xi_B})\cosh(\sqrt{\frac{-2V_1(l_0)}{\sigma}}
\frac{x}{\xi_B})\Big)\Big]& \textrm{for $x\leq R$}\\
\log\Big[\frac{2b}{\tau_2}
\Big(-1-e^{\frac{\tau_2}{\sqrt{2\sigma b}}(x-C)/\xi_B}\Big)\Big]  & 
\textrm{for $x>R$; }
\end{array} \right.
\eeq
whilst for case 2) we have
\beq
\label{l2}
\frac{\tilde{l}(x,l_0)}{\xi_B} = \left\{ \begin{array}{ll}
\log\Big[\frac{1}{2V_1(l_0)}\Big(\tau_1
+(\tau_1+2be^{-l_0/\xi_B})\cos(\sqrt{\frac{2V_1(l_0)}{\sigma}}
\frac{x}{\xi_B})\Big)\Big]& \textrm{for $x\leq R$}\\
\log\Big[\frac{2b}{\tau_2}\Big(-1-e^{\frac{\tau_2}
{\sqrt{2\sigma b}}(x-C)/\xi_B}\Big)\Big]  & \textrm{for $x>R$.}
\end{array} \right.
\eeq
The above expressions are valid for $x\geq 0$ and negative $x$-values 
are reached via  $\tilde l (-x,l_0)=\tilde l (x,l_0)$. 
The constant $C$ is in each case determined by the continuity condition of 
$\tilde l (x,l_0)$ at $x=R$. Note, that 
$\tilde{l}(x=R^-;l_0)$ is uniquely given by $l_0$. We require that $l_0$ 
is such that 
\beq
\label{warunek}
\tilde{l}(x=R^-,l_0)\geq l_{\pi2}\quad.
\eeq
 The necessity to make this additional though rather natural assumption is 
 a consequence of the 
discontinuous form of the interfacial potential $V(l,x)$, see Eq.(\ref{V}), 
which allowed us to solve 
Eq.(\ref{EL}) for $x<R$ and $x>R$ independently. The profiles 
$\tilde{l}(x,l_0)$ are decreasing functions of 
$x$, concave for $x<R$ and convex for $x>R$. 
\begin{figure}[h]
\begin{center}
\includegraphics[width=0.55\textwidth]{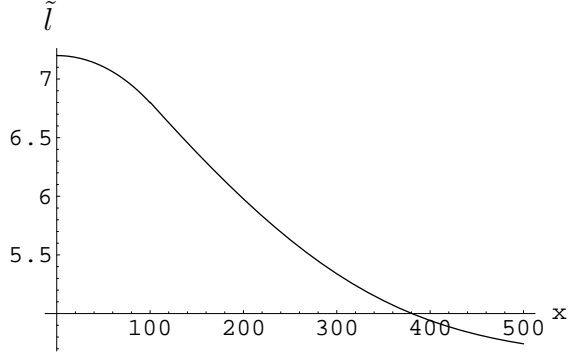}
\caption{An example of constrained equilibrium droplet profile $\tilde{l}(x,l_0)$. 
The plot parametres were chosen as follows: $2b/\sigma=1$, 
$\tau_1/\sigma=0.1$, $\tau_2/\sigma=-0.01$. The lengths $x$, $l_0=7.2$, 
$R=100$ are expressed in units of $\xi_B$. The profile's first derivative 
is discontinuous at $x=R$ (not visible in this figure's scale). 
The magnitude of the derivative discontinuity decreases to zero for 
$l_0\to\bar{l}_0$.} 
\end{center}
\end{figure}

At this stage one may calculate the constrained line contribution to 
the free energy per unit length in the 
$y$-direction as function of $l_0$ and $R$; it  is defined as
\beq
\label{H(l_0,R)}
H(l_0,R)=\int_0^{\infty} dx\Bigg[\frac{\sigma}{2}
\Big(\frac{d\tilde{l}(x,l_0)}{dx}\Big)^2+V(x,\tilde{l}) - 
V_1(l_{\pi 1})\Theta(R-x)-V_2(l_{\pi 2})\Theta(x-R)\Bigg].
\eeq           
Substituting expression (\ref{l1}) for case 1) and (\ref{l2}) 
for case 2) we obtain the following formulae  

$$
\frac{H(l_0,R)}{\xi_B}=\sqrt{2\sigma b}\Big(\frac{\tau_2}{2b}+
\frac{2V_1(l_0)}{\tau_1+(\tau_1+2be^{-l_0/\xi_B})\cosh(\alpha)}\Big)-
V_1(l_0)\frac{R}{\xi_B}+$$
\beq
\label{Naplin1}
+2\tau_1\sqrt{\frac{\sigma}{2b}}\textrm{arth}
\Big(\sqrt{\frac{b}{-V_1(l_0)}}e^{-l_0/\xi_B}\textrm{tgh}
\frac{\alpha}{2}\Big)-\sqrt{-2\sigma V_1(l_0)}\frac{\sinh\alpha}
{\frac{\tau_1}{\tau_1+2be^{-l_0/\xi_B}}+\cosh\alpha}+
\eeq 
$$
-\sqrt{\frac{\sigma}{2b}}\tau_2\log\Big[\frac{-\tau_2}{4bV_1(l_0)}
\Big(\tau_1+(\tau_1+2be^{-l_0/\xi_B})\cosh(\alpha)\Big)\Big]+
\frac{\tau_1^2}{4b}\frac{R}{\xi_B}\Theta(-\tau_1)
$$
\\
for case 1), and 

$$
\frac{H(l_0,R)}{\xi_B}=\sqrt{2\sigma b}\Big(\frac{\tau_2}{2b}+
\frac{2V_1(l_0)}{\tau_1+(\tau_1+2be^{-l_0/\xi_B})\cos\alpha}\Big)-
V_1(l_0)\frac{R}{\xi_B}+
$$
\beq
\label{Naplin2}
+2\tau_1\sqrt{\frac{\sigma}{2b}}\textrm{arth}\Big(\sqrt{\frac{b}
{V_1(l_0)}}e^{-l_0/\xi_B}\textrm{tg}\frac{\alpha}{2}\Big)+
\sqrt{2\sigma V_1(l_0)}\frac{\sin\alpha}{\frac{\tau_1}
{\tau_1+2be^{-l_0/\xi_B}}+\cos\alpha}+
\eeq 
$$
-\sqrt{\frac{\sigma}{2b}}\tau_2\log\Big[\frac{-\tau_2}{4bV_1(l_0)}
\Big(\tau_1+(\tau_1+2be^{-l_0/\xi_B})\cos\alpha\Big)\Big]
  +\frac{\tau_1^2}{4b}\frac{R}{\xi_B}\Theta(-\tau_1)
$$
\\
for case 2), where $\alpha=\sqrt{|\frac{2V_1(l_0)}{\sigma}|}\frac{R}{\xi_B}$. 
Note that the set of physically relevant values of $l_0$ is limited 
by the condition (\ref{warunek}), which means that $l_0>l_{0 min}$, 
where $l_{0 min}$ is given by $\tilde{l}(x=R,l_{0 min})=l_{\pi 2}$. 
The inhomogeneity width $R$ is arbitrary, nevertheless in further 
calculations we shall assume $\xi_{||i}\ll R$, where $\xi_{||i}$ 
denotes the mean-field interfacial correlation length corresponding 
to substrate of type $i=1,2$, $\xi_{|| i}^{-2}=V_i''(l_{\pi i})/\sigma$.  

The formulae (\ref{l1}), (\ref{l2}), (\ref{Naplin1}), and (\ref{Naplin2}) 
form the starting point for the subsequent 
analysis aiming at the determination of the droplets' height, 
line tension, and the contact angles.
The line tension $\eta$ as function of {$\tau_i$} and $R$ is 
given as the minimum of $H(l_0,R)$ with respect to $l_0$. 
The corresponding value of drolet's height is denoted by $\bar{l}_{0}$. Consequently, 
the equilibrium interfacial shape  
$\bar{l}(x) = \tilde{l}(x,\bar{l}_{0})$, and $\eta = H(\bar{l}_0,R)$.
 
\section{The equilibrium droplet height}
\renewcommand{\theequation}{3.\arabic{equation}} 
\setcounter{equation}{0}
\vspace*{0.5cm}

The scaling behaviour of the equilibrium droplet height 
$\bar{l}_0$ for $R\to\infty$ and 
$\tau_1>0$,  was 
obtained by Parry \textsl{et al} using conformal properties of 
equation (\ref{EL}) in the case 
$b=0$ \cite{Conf}. However, 
this technique fails when the term $\sim e^{-2l/\xi_B}$ in 
$V_1(l)$ must be taken into account. As we show below, the term 
neglected in \cite{Conf} contributes to the corrections to the leading 
behaviour of $\bar{l_0}$ which are 
essential to determine the character of the asymptotic expansion 
of line tension $\eta$ for large 
$R$. In the following subsections we 
expand the functions $\frac{\partial H(l_0,R)}{\partial l_0}$ around the 
properly chosen values 
$l_0^{(0)}$ of $l_0$ 
for different temperature ranges. We assume 
$\bar l_0=l_0^{(0)}+\delta l_0$, $\delta l_0\ll l_0^{(0)}$, and 
solve the appropriate equation for $\delta l_0$, thus determining 
$\bar{l}_0$ up to terms necessary 
for the calculation of the leading correction to $\eta$ due to finite values of $R$.

\subsection{The $\tau_1>0$ regime}
In this case the numerical analysis of the constrained line tension 
$H(l_0,R)$ given by 
Eq.(\ref{Naplin2}) shows that the equilibrium droplet height fulfils 
the  condition 
$\alpha (\bar{l}_0(R))\to \pi^-$, as $R\to\infty$. The range of 
physically relevant values of $l_0$ is limited by the condition 
(\ref{warunek}). It follows  that only 
$l_0>l_{0 min}$ should be considered, where $l_{0 min}$ is defined by 
$\tilde{l}(x=R,l_{0 min})=l_{\pi 2}$. 
We note that $H(l_0,R)$ diverges at $l_0 =l_0^*<l_{0 min}$ and one 
can check that asymptotically 
$l_0^*\to l_{0 min}$ and simultanously $\bar{l}_0\to l_0^*$, for 
$R\to\infty$. One may not perform 
the expansion of $\frac{\partial H(l_0,R)}{\partial l_0}$ around 
the value of $l_0$ which 
corresponds to $\alpha=\pi$, as this value is smaller than $l_0^{*}$ 
for finite $R$ (see Fig.4).       
     
\begin{figure}[h]
\begin{center}
\includegraphics[width=0.5\textwidth]{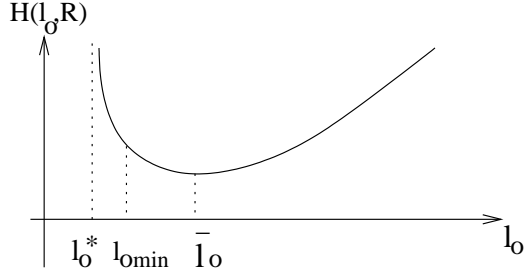}
\caption{A sketch of $H(l_0,R)$ for fixed $R$. The physically meaningful 
values of droplet height fulfil $l_0>l_{0 min}$, 
where $l_{0min}$ is determined by the condition 
$\tilde{l}(x=R,l_{0 min})=l_{\pi 2}$. The function $H(l_0,R)$ diverges at 
$l_0^{*}<l_{0 min}$.}  
\end{center}
\end{figure}
Instead, one expands $\frac{\partial H(l_0,R)}{\partial l_0}$ around 
$l_0^{(0)}= l_0^{*}$, which fulfils 
\beq
\label{warunek2}
\tau_1+(\tau_1+2be^{-l_0^{(0)}/\xi_B})\cos[\alpha(l_0^{(0)},R)]=0.
\eeq
The solution of Eq.(\ref{warunek2}) takes - for 
$\frac{\sigma\xi_B}{\tau_1 R}\ll 1$ - the following form 
\beq
e^{-l_0^{(0)}/\xi_B}=\frac{\sigma\pi^2\xi_B^2}{2\tau_1 
R^2}\Big(1-2\frac{\sqrt{2\sigma b}}{\tau_1}\frac{\xi_B}{R}+
\frac{1}{2}(6-\pi^2/2)\frac{2\sigma b}{\tau_1^2}\frac{\xi_B^2}{R^2}+
\mathcal{O}(\frac{\sigma\xi_B}{\tau_1R})^3\Big)\quad.
\eeq
Similarly, by expanding $\frac{\partial H(l_0,R)}{\partial l_0}$ 
around the above value $l_0^{(0)}$, we solve the equation 
$\frac{\partial H(l_0,R)}{\partial l_0}=0$ up to terms of 
the order $(\frac{\xi_B\sigma}{\tau_1 R})^4$.
In this way one obtains the equilibrium droplet height 
\beq
\label{l01}
e^{-\bar{l}_0/\xi_B}=\frac{\sigma\pi^2\xi_B^2}{2\tau_1 R^2}
\Big(1+\frac{\kappa_1}{R}+\frac{\kappa_2}{R^2}+\mathcal{O}
((\frac{\sigma\xi_B}{\tau_1R})^3 )\Big),
\eeq
where $\kappa_1=\xi_B\sqrt{8\sigma b}\,\frac{2\tau_1-\tau_2}{\tau_1\tau_2}$, 
$\kappa_2=8\sigma b\xi_B^2\Big[\frac{6-\frac{\pi^2}{2}}{8\tau_1^2}
+\frac{2}{\tau_2^2}-\frac{2}{\tau_1\tau_2}-\frac{-2\tau_1^3
+5\tau_1^2\tau_2-2\tau_1\tau_2^2+\tau_2^3}{\tau_1\tau_2^2(2\tau_1^2
-3\tau_1\tau_2+\tau_2^2)}\Big]$. If in Eq.(\ref{l01}) one 
neglects the correction terms of the order 
$\sim(\frac{\sigma\xi_B}{\tau_1R})^3 $ and 
$\sim(\frac{\sigma\xi_B}{\tau_1R})^4 $, 
then one recovers the result obtained 
by Parry \textsl{et al} via conformal transformation \cite{Conf}. 

The droplet height 
$l_0=\xi_B \left[\log\left(\frac{2\tau_1R^2}{\pi^2\sigma\xi_B^2}\right) 
- \frac{\kappa_1}{R}-\frac{\kappa_2-\frac{1}{2}\kappa_1^2}{R^2}+...\right]$  
exhibits logarythmic divergence for $R\to\infty$. The first correction 
($\sim 1/R$) 
is positive. Both $\kappa_1$ and $\kappa_2$ depend on $b$ and 
thus cannot be obtained within the 
conformal scheme \cite{Conf}. Note, that the asymptotic expansion was performed 
while keeping $\tau_i$ ($i=1,2$) constant. 
Taking the limits $\xi_B/R\to 0$ and $\tau_i/\sigma\to 0$ are 
non-commutative operations and - consequently - 
the above expansion (\ref{l01}) is valid only provided 
$\frac{\xi_B\sigma}{R\tau_1}\ll1$. This remark remains also 
relevant for the cases $\tau_1=0$ and $\tau_1<0$ discussed below.

\subsection{The $\tau_1=0$ regime}
The plot of the function $H(l_0,R)$ for $\tau_1=0$ case is 
qualitatively similar to the $\tau_1>0$ case, see Fig.4. 
However, in the present case the quantity $\bar l_0$ 
corresponds to $\alpha (\bar l_0(R))\to\frac{\pi}{2}^-$ for 
$R\to\infty$. Following, we may expand $H(l_0,R)$ around 
$l_0^{(0)}$ which fulfils the condition 
$\alpha (l_0^{(0)},R)=\pi/2$. After solving the equation 
$\frac{\partial H(l_0,R)}{\partial l_0}=0$ we obtain
\beq
e^{-\bar{l}_0/\xi_B}=\sqrt{\frac{\sigma}{2b}}\frac{\pi}{2}
\frac{\xi_B}{R}\Big(1+\frac{\gamma_1}{R}+
\frac{\gamma_2}{R^2}+\mathcal{O}((\frac{\xi_{||2}}{R})^3)\Big)\quad,
\eeq
where $\gamma_1=-2\xi_{||2}$, $\gamma_2=4\xi_{||2}^2$. 
The explicit expression for $\bar{l}_0$ has the form 
$\bar{l}_0=\xi_B[\log\sqrt{\frac{2b}{\sigma}}\frac{2}{\pi}\frac{R}{\xi_B}
+\frac{2\xi_{||2}}{R}-\frac{2\xi_{||2}^{2}}{R^2}+...]$. 
The leading correction term, alike case $\tau_1>0$, is positive.

\subsection{The $\tau_1<0$ regime}
Our first observation concerning the case $\tau_1<0$ is that in the 
limit $R\to\infty$, the 
equilibrium droplet height $\bar{l}_0(R)$ converges to a finite value 
$l_{\pi 1}$ corresponding to the 
minimum of the potential $V_1(l)$. Indeed,  as the inhomogeneity size 
becomes infinite, the height of the 
drop at its centre 
is fully determined by the properties of substrate type 1. The quantity 
$l_0^{(0)}=l_{\pi 1}$ is therefore a 
natural choice of the value around which 
$\frac{\partial H(l_0,R)}{\partial l_0}$ can be expanded for large 
$R$. Note that unlike the former cases $\tau_1\geq 0$, the 
effective potential $V_1(l_0)$ becomes 
negative close to $l_0=\bar{l}_0(R)\simeq l_{\pi 1}$ and 
so the formula in Eq.(\ref{Naplin1}) should be used instead of 
(\ref{Naplin2}). Solving $\frac{\partial H(l_0,R)}{\partial l_0}=0$ 
up to the relevant order leads to 
\beq
\bar{l}_0=l_{\pi 1}+\delta_1 e^{-R/\xi_{||1}} +\delta_2 e^{-2R/\xi_{|| 1}}
+\mathcal{O}(e^{-3R/\xi_{|| 1}})\quad,
\eeq
where  $\delta_1=2\frac{\tau_1-\tau_2}{\tau_1+\tau_2}$, and 
$\delta_2=\frac{1}{2}\delta_1^2$. The leading
correction is negative and equal $0$ for the special case 
$\tau_1=\tau_2$ as is expected on physical ground. 

To sum up the results obtained for the morphology of the 
liquid drop we note that the form of the 
$\bar l_0$ expansion for large $R$ depends crucially on the value of the 
reduced temperature 
$\tau_1$. For $\tau_1\geq 0$ which corresponds to complete wetting of 
type 1 substrate one has 
corrections proportional to powers of $1/R$ while for $\tau_1<0$ we obtain 
corrections proportional to powers of $e^{-R/\xi_{||1}}$.    

\section{The line tension} 
\renewcommand{\theequation}{4.\arabic{equation}} 
\setcounter{equation}{0}
\vspace*{0.5cm}

The  expressions for the line tension $\eta(R)=H(\bar{l}_0,R)$ are 
obtained by substituting the formulae 
for $\bar{l}_0$ into equations (\ref{Naplin1}),  and (\ref{Naplin2}). 
The calculations are rather cumbersome 
and we refrain from qouting the expressions for $\eta$ in terms of 
the parametres of the expansion 
of $\bar{l}_0(R)$ around $R=\infty$. We only note that the leading 
$R$-independent term in $\eta$ 
may be expressed in terms of the parametres $\kappa_1$, $\gamma_1$ or 
$\delta_1$ for all 
cases $\tau_1>0$, $\tau_1=0$, $\tau_1<0$, respectively. The next-to 
leading corrections in 
$\bar{l}_0$ modify $\eta$ at the level of leading $R$-dependent terms.    

We obtain the following expressions:
   
\begin{eqnarray}
\label{Eta1}
\eta_>(R,\tau_1,\tau_2) =  
\xi_B\sqrt{\frac{\sigma}{2b}}\Big[\tau_1\log\Big(\frac{\tau_1-
\tau_2}{\tau_1}\Big)-\tau_2\log\Big(\frac{2(\tau_2-\tau_1)}
{\tau_2}\Big)\Big]-  \nonumber \\\sigma\xi_B \Big[\frac{\pi^2}{2}\frac{\xi_B}{R}
+\mathcal{O}((\frac{\xi_B}{R})^2)\Big]\quad,
\end{eqnarray}

for $\tau_1>0$, 

\beq
\label{Eta2}
\eta_0(R,\tau_1,\tau_2) = 
-\xi_B\tau_2\sqrt{\frac{\sigma}{2b}}\log (2)-
\sigma\xi_B \Big[\frac{\pi^2}{8}\frac{\xi_B}{R}+
\mathcal{O}\Big((\frac{\xi_B}{R})^2\Big)\Big]\quad,
\eeq
for $\tau_1=0$, 

and finally
\begin{eqnarray}
\label{Eta3}
\eta_<(R,\tau_1,\tau_2) = \xi_B\sqrt{\frac{\sigma}{2b}}
\Big[\tau_1\log\Big(\frac{\tau_1+\tau_2}{2\tau_1}\Big)+
\tau_2\log\Big(\frac{\tau_1+\tau_2}{2\tau_2}\Big)\Big] + \nonumber \\
\xi_B\tau_1\sqrt{\frac{\sigma}{2b}}
\Big[- 2\frac{\tau_1}{\tau_1+\tau_2}e^{-R/\xi_{||1}} + 
\mathcal{O}\Big(\frac{R}{\xi_B}e^{-2R/\xi_{||1}}\Big)\Big]\quad,
\end{eqnarray}
for $\tau_1<0$. 

The dominant, $R$-independent terms in the above formulae may be obtained 
independently by considering an inhomogeneous substrate consisting of two 
homogeneous half-planes characterized by $\tau_1$, $\tau_2$, 
meeting at $x=0$. After calculating the corresponding excess linear free 
energy one obtains the dominant, $R$-independent term. It is a non-negative 
function of  
$\tau_1$ and $\tau_2$, which vanishes for $\tau_1 = \tau_2$, see Fig.5.
\begin{figure}[h]
\begin{center}
\includegraphics[width=0.55\textwidth]{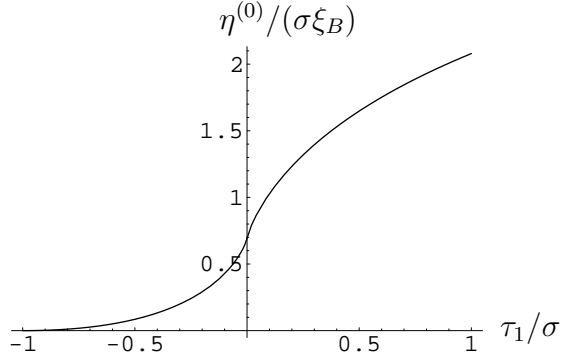}
\caption{Plot of the leading $R$-independent contribution to line tension 
denoted by $\eta^{(0)}$ 
as function of $\tau_1$ at fixed $\tau_2$. The plot parametres were chosen 
as follows: $\sigma/2b=-\tau_2/2b=1$. 
Note that for $\tau_1=\tau_2$ one has $\eta^{(0)}=0$ as expected on 
physical ground. The range of $\tau_{1}$ is restricted to 
$\tau_{1}>\tau_{2}$ in accordance with the analysis in the text.} 
\end{center}
\end{figure}

The central observation concerning the above formulae is the change 
in the character of the corrections 
to the "free" line tension $\eta^{(0)}$ as the temperature $T_{W1}$ 
is crossed. The interfacial correlation length 
$\xi_{||1}=\frac{\sqrt{2\sigma b}}{|\tau_1|}$ sets the length 
scale controlling the next-to-leading terms in 
$\eta_<$, i.e., for $\tau_{1}<0$. Its divergence at $\tau_1=0$ 
is accompanied by the appearance of the algebraic terms in the 
expansion of $\eta$ about $1/R=0$ for $\tau_{1}\ge0$ . The finite 
$R$ corrections for $\tau_1>0$ 
are universal (at least up to $\sim 1/R$ order) in the sense that 
they do not depend on the substrates' 
characteristics and are fully determined by $\sigma$. Thus the 
divergence of $\xi_{||1}$ at $\tau_1=0$ is 
accompanied by the change of the expansion parameter of $\eta$ 
from $e^{-R/\xi_{||1}}$ to $\xi_B/R$.

The corrections are positive for $\tau_1<0$ and negative for $\tau_1\geq 0$.  
They may be interpreted as the 
asymptotic term of an effective interaction potential between two regions 
of inhomogeneity of 
$\bar{l}(x)$, corresponding to $x\sim R$ and $x\sim -R$.   

\section{Remarks on the contact angles}
\renewcommand{\theequation}{5.\arabic{equation}} 
\setcounter{equation}{0}
\vspace*{0.5cm}
 
At the macroscopic level the  contact angle of a liquid drop placed 
on a planar solid surface is related to 
the  interfacial tensions between the relevant phases. Accordingly, 
it can be evaluated from the Young's equation \cite{RowWid}. 
At the mesoscopic level the 3-phase contact line is not defined 
unambigously and consequently the geometric 
construction of the contact angle is not unique. 
Moreover, generalizations of Young's equation to the case of 
chemically inhomogeneous substrates assume 
implicitely that the typical size of boundary regions between 
homogeneous domains is much larger than bulk correlation length 
$\xi_B$ (see Ref.\cite{Sw1} for detailed discussion), which is opposite 
to the case considered in this paper. 

There are at least two approaches towards the geometric definition 
of the contact angle for the 
system considered in this paper. The  first definition exploits the 
interfacial profile's characteristics in the region of 
inhomogeneity (i.e., for $|x|\approx R$) while the second probes the  
profile's features at the centre of the 
droplet (i.e., $x\approx 0$). 

Within the first approach the equilibrium contact angle is 
defined via the profile's slope at the 
three phase contact line situated at a choosen position $x_k$. 
A natural - though not unique - choice for 
the considered system is $x_k=R$. Following, the contact angle 
$\theta$ is obtained from  
\beq
\left.\tan \theta=-\frac{d \bar{l}(x)}{dx}\right|_{x=R}\quad .
\eeq
In the second approach one fits the circle 
$y(x)=\sqrt{\mathcal{R}^2-x^2}+y_0$, tangent to 
$\bar{l}(x)$ at $x=0$, and of the same curvature at $x=0$ as 
that of $\bar{l}(x)$:  
$\frac{d^2y(x)}{dx^2}|_{x=0}=-\frac{1}{\mathcal{R}}=
\frac{d^2\bar{l}(x)}{dx^2}|_{x=0}$. 
The position of the contact line is then defined as the line of 
intersection of $y(x)$ with 
the profile's asymptote at $l_{\pi 2 }$, see Fig.6. 
\begin{figure}[h]
\begin{center}
\includegraphics[width=0.55\textwidth]{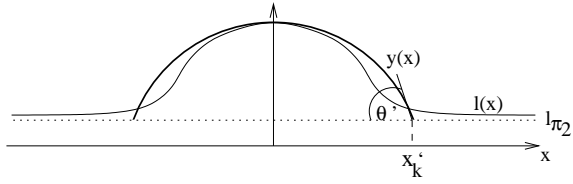}
\caption{Phenomenological construction of the contact line 
and the contact angle. 
A circle $y(x)$ fitted to the equilibrium droplet profile at $x=0$ intersects 
the asymptote $l_{\pi 2}$ at $x_k'$ defining the three phase contact line 
position. The contact angle $\theta '$ is defined as the corresponding angle 
of intersection.} 
\end{center}
\end{figure}

Within the present analytical approach the above two constructions 
can be achieved only in the case 
of large inhomogeneity width $R/\xi_{||1}\gg 1$. For the first 
choice one obtains 
\beq
\label{theta1}
\theta\simeq\left\{ \begin{array}{ll}
\frac{1}{2}(\theta_2-\theta_1) & \textrm{for $\tau_1\leq0$}\\
\frac{1}{2}\frac{\theta_2^2}{\frac{\tau_1}{\sqrt{2\sigma b}}+\theta_2}
  & \textrm{for $\tau_1>0$}
\end{array} \right. \quad, 
\eeq
where $\theta_1$ and $\theta_2$ are contact angles for homogeneous 
substrates of type $1$ and $2$, respectively, and 
for small, negative $\tau_{i}$ one has  $\theta_i\simeq |\tau_i|/\sqrt{2\sigma b}$, 
$i=1,2$. \\
It is obvious on physical ground that the second construction leads for 
$R/\xi_{||1}\gg 1$ to the vanishing contact 
angle. This is confirmed by the analytical expressions

\beq
\label{theta2}
\theta'\simeq\left\{ \begin{array}{ll}
2|\tau_1|\sqrt{(l_{\pi 1}-l_{\pi 2})\frac{\tau_2-\tau_1}{\tau_1+
\tau_2}\frac{1}{2\sigma b}}e^{-\frac{1}{2}R/\xi_{||1}} & 
\textrm{for $\tau_1<0$}\\
\frac{\pi}{\sqrt{2}}\frac{\sqrt{\log[-(2/(\sigma b))^{1/2}
\tau_2 R/\pi]}}{R} &\textrm{for $\tau_1=0$}\\
\pi\frac{\sqrt{\log[\frac{\tau_1\tau_2 R^2}{-\pi^2\sigma b}]}}{R}
  & \textrm{for $\tau_1>0$}
\end{array} \right.\quad,
\eeq  
which display change of the convergence type to the value $\theta'(R=\infty)=0$ 
from exponential for $\tau_1<0$ to 
$\frac{\sqrt{\log(R)}}{R}$ for $\tau_1\geq 0$. In other words, the 
discrepancy between our predictions for $\theta$ and $\theta'$ may be attributed 
to different behaviour of the corresponding positions  $x_k$ and $x_k'$ of 
the contact line for $R \rightarrow \infty$. 
In particular, the difference $x_k-x_k'$ is of the order $e^{R/(2\xi_{||1})}$ 
for $\tau_1<0$ and $\sqrt{\log R}$ for $\tau_1\geq 0$. 

\section{Summary}
\renewcommand{\theequation}{6.\arabic{equation}} 
\setcounter{equation}{0}
\vspace*{0.5cm}
This article is concerned with morphology and line tension of 
equilibrium droplets placed on a planar substrate with a chemical, 
stripe-shaped inhomogeneity. 

Applying the effective Hamiltonian approach, we first constructed a 
set of constrained equilibrium states parametrized by the droplet's 
height. The global minimum was then extracted from this set 
using perturbative expansion valid in the limit of large droplet 
radii. The line tension expressions were evaluated and the character 
of the size-dependent 
correction shown to exhibit crossover from exponential to algebraic 
decay for increasing  inhomogeneity's 
width, occuring at the strip's wetting temperature. This corresponds 
to divergence of the interfacial 
correlation length characterizing the wetted part of the surface, 
which acts as the length scale 
controlling line tension for low temperatures. \\
In addition, on the basis of the obtained equilibrium liquid-vapour 
profiles $\bar{l}(x)$ we calculated the 
droplet contact angle on the heterogeneous substrate defined via 
tangents of $\bar{l}(x)$ evaluated at $x=R$. 

The main conclusions are restricted to the case of short-ranged 
forces and refer to the asymptotic regime 
$R/\xi_{||1}\to\infty$, and the thermodynamic state close to the 
fluid two-phase coexistence ($\mu=\mu_{sat}^-$). 
Our predictions  are as follows:
\begin{itemize}
\item
The line tension $\eta$ related to the  substrate's inhomogeneity is 
a function of the stripe's width $R$. For large $R$ the character of 
the leading $R$-dependent term depends on the temperature. At low 
temperatures, i.e., for $T<T_{W1}$ the interfacial correlation length  
$\xi_{||1}$ represents the characteristic length scale controlling the 
$R$-dependent contributions to line tension, which are of the type 
$e^{-R/\xi_{||1}}$. For $T\geq T_{W1}$ the correlation length 
$\xi_{||1}$ is infinite and the $R$-dependent contributions to 
$\eta$ become algebraic and universal, i.e. independent of 
substrates' properties, and fully determined by the liquid-vapour 
interfacial tension $\sigma$.
\item
The $R$-dependent contributions to $\eta$ may be interpreted as the  
effective interaction potential between the interfacial profile 
inhomogeneity regions 
corresponding to droplet boundaries at $x\approx R$ and $x\approx -R$. 
At $T=T_{W1}$ the leading $R$-dependent correction to $\eta$ changes 
sign from positive in the low temperature regime $T<T_{W1}$ to 
negative for $T\geq T_{W1}$. This means that the interaction 
represented by these terms is 
repulsive as long as $T$ does not exceed $T_{W1}$ and attractive  
for $T>T_{W1}$. The interaction magnitude is governed by the 
inverse of the correlation length 
$1/\xi_{||1}=|\tau_1|/\sqrt{2\sigma b}$ for $T<T_{W1}$ and by 
the interfacial tension $\sigma$ in the case of $T\geq T_{W1}$.       
\item
In order to characterize the interfacial morphology in terms of 
the contact angle we analysed two definitions of the contact angle. 
They turn out to be highly sensitive to the choice of the three phase 
contact line.  A particular choice of this line positioned at the domain 
boundary leads to a contact angle expressible in terms of contact 
angles on homogeneous substrates. Another choice, inspired by 
experimental procedures and based on droplets' properties at their 
centres, leads - in the present context - to contact angles 
asymptotically (for $R\to\infty$) equal $0$ for all temperatures.   
\end{itemize} 
\vspace{1cm}
{\bf {Acknowledgment}} The support by the Committee for Scientific Research  
grant KBN 2P03B 008 23 is gratefully acknowledged.\\


\newpage
\begin{thebibliography}{99}
\bibitem{Hau} Hauge E 1992 \textsl{Phys. Rev. A} \textbf{46} 4994
\bibitem{R3} Rejmer K, Dietrich S and Napi\'{o}rkowski M 1999 
\textsl{Phys. Rev. E} 
\textbf{60} 4027
\bibitem{P1} Parry A, Rasc\'{o}n C and Wood A 1999 
\textsl{Phys. Rev. Lett.} 
\textbf{83} 5535
\bibitem{P2} Parry A, Rasc\'{o}n C and Wood A 2000 
\textsl{Phys. Rev. Lett.} \textbf{85} 345



\bibitem{Dr} Drelich J, Wilbur J, Miller J and Whitesides G 1996 
\textsl{Langmuir} 
\textbf{12} 1913
\bibitem{Urb} Urban D, Topolski K and De Coninck J. 1996 
\textsl{Phys. Rev. Lett.} 
\textbf{76} 4388
\bibitem{Sw1} Swain P and Lipowsky R 1998 \textsl{Langmuir} \textbf{14} 6772
\bibitem{D2} Bauer C and Dietrich S 1999 \textsl{Phys. Rev. E} 
\textbf{60} 6919
\bibitem{D1} Dietrich S and Bauer C 2000 \textsl{Phys. Rev. E} 
\textbf{62} 2428 
\bibitem{Le1} Lenz P, Bechinger C, Schafle C, Leiderer P and Lipowsky R 2001 
\textsl{Langmuir} \textbf{17} 7814


\bibitem{Le2} Lenz P and Lipowsky R 1998 
\textsl{Phys. Rev. Lett.} \textbf{80} 1920
\bibitem{Sw2} Swain P and Lipowsky R 2000 
\textsl{Europhys. Lett.} \textbf{80} 203


\bibitem{Hol} Ho\l yst R and Poniewierski A 1987 
\textsl{Phys. Rev. B} \textbf{36} 5628
\bibitem{Ind1} Indekeu J, Upton P and Yeomans J 1988 \textsl{Phys. Rev. Lett.} 
\textbf{61} 2221
\bibitem{N1} Netz R and Andelman D 1997 \textsl{Phys. Rev. E} 
\textbf{55} 687
\bibitem{R1} Rejmer K and Napi\'{o}rkowski M 1997 
\textsl{Z. Phys. B} \textbf{102} 101;\textsl{Phys. Rev. E} \textbf{62} 588 
\bibitem{Ras1} Rasc\'{o}n C and Parry A 2000 \textsl{J. Phys.:Cond. Matt.} 
\textbf{12} A369


\bibitem{RowWid} Rowlinson R and Widom B 1982 
\textsl{Molecular Theory of Capillarity}, Oxford 
University Press, New York  

\bibitem{Wi1} Widom B and Clarke A 1990 \textsl{Physica A} \textbf{168} 149
\bibitem{Ind2} Indekeu J 1992 \textsl{Physica A} \textbf{183} 439
\bibitem{Va} Varea C and Robledo A 1992 \textsl{Phys. Rev. A} \textbf{45} 2645
\bibitem{Ab} Abraham D, Latr\'emoli\'ere F, and Upton P 1993 \textsl{Phys. Rev. Lett.} 
\textbf{71} 404
\bibitem{Ind4} Indekeu J and Robledo A 1993 \textsl{Phys. Rev. E} \textbf{47} 4607
\bibitem{Ind3} Indekeu J 1994 \textsl{Int. J. Mod. Phys. B} \textbf{8} 309
\bibitem{K1} Koch W, Dietrich S and Napi\'{o}rkowski M 1995 
\textsl{Phys. Rev. E} 
\textbf{51} 3300
\bibitem{Ge} Getta T and Dietrich S 1998 \textsl{Phys. Rev. E} \textbf{57} 655
\bibitem{Pom2} Pompe T and Herminghaus S 1999 \textsl{Phys. Rev. E} \textbf{83} 3677
\bibitem{Wa} Wang J, Betelu S and Law B 2001 \textsl{Phys. Rev. E} \textbf{63} 031601
\bibitem{Pom} Pompe T 2002 \textsl{Phys. Rev. Lett.} \textbf{89} 076102




\bibitem{Conf} Parry A, Macdonald E Rasc\'{o}n C 2001 \textsl{J. Phys.: Cond. Matt} 
\textbf{13} 383 
\bibitem{Bau1} Bauer C, Dietrich S and Parry A 1999 \textsl{Europhys. Lett.} 
\textbf{47} 474
\bibitem{Bau2} Bauer C and Dietrich S 1999 \textsl{Eur. Phys. J. B} \textbf{10}  767 
\bibitem{Dobbs} Dobbs H 1999 \textsl{Int. J. Mod. Phys. B} \textbf{13} 3255  


\end{thebibliography}
\end{document}